\documentclass[journal ]{new-aiaa}
\usepackage[utf8]{inputenc}
\usepackage{textcomp}

\usepackage{graphicx}
\usepackage{amsmath}
\usepackage[version=4]{mhchem}
\usepackage{siunitx}
\usepackage{longtable,tabularx}
\newcommand{\be}{\begin{equation}}
\newcommand{\ee}{\end{equation}}
\newcommand{\ben}{\begin{eqnarray}}
\newcommand{\een}{\end{eqnarray}}

\newcommand{\br}{\mbox{\boldmath $r$}}

\newcommand{\bv}{\mbox{\boldmath $v$}}

\newcommand{\bff}{\mbox{\boldmath $f$}}
\newcommand{\bx}{\mbox{\boldmath $x$}}
\newcommand{\bu}{\mbox{\boldmath $u$}}

\usepackage{setspace}
\usepackage{graphicx}
\usepackage{epstopdf}
\usepackage{dcolumn}
\usepackage{bm}
\usepackage{verbatim}
\doublespace{}

\begin{document}
\title{A mass-optimized gravity tractor for asteroid deflection}
\author{Yohannes Ketema\footnote{Associate Professor, Dept. of Aerospace Engineering and Mechanics, 107 Akerman Hall, 110 Union Street SE, Associate Fellow AIAA}\\University of Minnesota, Minneapolis, MN 55455}
\maketitle

\begin{abstract}
A method for asteroid deflection that makes use of a spacecraft moving
back and forth on a segment of a Keplerian orbit about the asteroid is
studied with the aim of optimizing the initial gross mass of the
spacecraft. The corresponding optimization problem is formulated as a 
discrete nonlinear optimal control problem where the parameters of the orbit
segment are the control variables. A hypothetical asteroid deflection problem is solved
numerically using the method of dynamic programming, and it is shown that a
gravity tractor can be obtained that is 
significantly more efficient in terms of deflection attained
per unit mass of the spacecraft, compared to similar gravity tractors in the literature.
\end{abstract}
\section*{Nomenclature}
\begin{tabular}{ l c l}
 $m_a$&=&mass of asteroid \\
$m_c$&=&mass of spacecraft\\
$\mu_a$&=&gravitational parameter for asteroid\\
$\mu_c$&=&gravitational parameter for spacecraft\\
$r_a$&=&radius of asteroid\\
$r$&=&distance to spacecraft from asteroid center \\
$\alpha$&=&ratio of spacecraft distance to asteroid radius\\
$G$&=&universal gravitational constant\\
$\theta$&=&true anomaly on spacecraft orbit about asteroid\\
$\theta_b$&=&bounding angle on spacecraft orbit about asteroid \\
$\Delta v$&=&impulsive velocity change\\
$I$&=&impulse imparted to asteroid \\
$h$&=&specific angular momentum for spacecraft orbit about asteroid\\
$r_p$&=&periapsis distance for spacecraft orbit about asteroid\\
$r_{pm}$&=&smallest allowable periapsis distance for spacecraft orbit about asteroid\\
$e$&=&eccentricity of spacecraft orbit about asteroid\\
$\gamma$&=&flight path angle for spacecraft orbit about asteroid\\
$\varphi$&=&plume half angle\\
$\Delta t$&=&time of flight on spacecraft orbit segment \\
\end{tabular}

\noindent
\begin{tabular}{ l c l}
$t_e$&=&projected time of Earth-asteroid encounter\\
$v_a$&=&orbital speed of asteroid\\
$a_a$&=&semi-major axis of asteroid's orbit\\
$\zeta$&=&asteroid deflection measured at time of encounter\\
$r_0$&=&distance to apsis of gravity tractor orbit segment from asteroid center\\
$\alpha$&=&inverse of semi-major axis of gravity tractor orbit segment\\
$\chi$&=&universal variable for the time of flight on the gravity tractor orbit segment
\end{tabular}

\vspace{.75in}
\section{Introduction}
The prospect of Earth being impacted by an asteroid with
potentially dire consequences to life on the planet calls for measures towards
preventing or mitigating such a disaster. Thus several studies in the literature
have been concerned with methods of asteroid deflection, i.e. altering the trajectories of asteroids
by a required amount in order to avert a projected collision with Earth.

%

Three main approaches have been proposed with regards to asteroid deflection. These
are kinetic
impactors  \cite{ahrens,mcinnes,dachwald,bernd,izzo1,asfaw}, nuclear
interceptors \cite{sanchez, ahrens,dearborn}, and gravity tractors
\cite{lu-love,gehler,wei,olympio,fahn,foster,ketema,mazanek}.  
Each method has
characteristics that may make it suitable for the deflection of 
certain types and sizes of asteroids and possibly not for others.  For
example, kinetic impactors may be used with the aim of imparting an
impulse to an asteroid to perturb its orbit without breaking it up
into smaller pieces, thereby avoiding the creation of several objects
that may impact Earth. In such a case, it is important that the asteroid be
of a rocky type that will not easily disintegrate.

On the other hand, gravity tractors do not involve any direct contact
with the asteroid. For this reason, they may be a viable approach for
asteroid deflection when the asteroid is likely to break up upon
impact. The force that gravity tractors can exert on an asteroid diminishes by
the inverse square law. Thus, gravity tractors would generally be most
effective on smaller asteroids \cite{wei}.



Various forms of gravity tractors have been studied in the
literature including the stationary gravity tractor \cite{lu-love}
with canted thrusters in order to avoid impingement of the thruster
plume on the asteroid; the displaced orbit (halo orbit) gravity
tractor that gives a better thrust/fuel-mass ratio \cite{mcinnes2,mcinnes3};
and gravity tractors using solar sails \cite{wei,gong}. In addition, the use
of multiple gravity tractors in order to enhance the overall force on the asteroid
has been studied in \cite{foster}. The navigation problem for such formations is studied in \cite{yang}.
 
The gravity tractor that is considered in this
paper is based on the method introduced in \cite{ketema} and, as will
be described further in the body of this paper, consists
of a spacecraft that flies back and forth in a type of ``restricted
Keplerian motion'' on an orbit segment about the asteroid. This
gravity tractor can exert a significantly
larger average force on the asteroid for a given spacecraft mass, while
consuming less fuel, than for example a stationary gravity tractor
with canted thrusters \cite{lu-love}, or a displaced orbit gravity
tractor \cite{mcinnes2}.

The goal of this paper is to further improve upon the efficiency
of the gravity tractor in (\cite{ketema}). This is done by optimizing 
the gross initial mass of the spacecraft for a given amount of
deflection by the projected date of encounter. The optimization takes into
account variables that define the characteristics of the restricted
Keplerian orbit segment as well as constraints that the orbit segment
must satisfy so that a) the spacecraft does not impact the asteroid,
b) there is no thruster plume impingement on the asteroid, and c)
the time of flight on the orbit segment is sufficiently large for
a realistic mission such that the spacecraft can properly orient and fire
its thrusters at the rate of succession specified. In a numerical
example consisting of a  hypothetical asteroid
deflection problem, it is shown that the optimization results in 
a decrease in the initial gross mass of the spacecraft that is 
in the order of 20\% compared to a similar problem in \cite{ketema}
where no optimization was done. 

Lastly, while the main focus of the work in
this paper pertains to asteroid deflection in the case of a projected
collision with Earth, the methods described would be applicable to related problems of asteroid redirection
missions (ARM). In addition, the method described is in essence a
way of "hovering" over an asteroid by remaining within given bounds
on a Keplerian orbit while minimizing the required amount of fuel, and may therefore have
broader applications, see e.g. \cite{broschart}.

The rest of this paper is organized as follows. In Section 2, a short
description of the type of gravity tractor considered, i.e. a gravity
tractor in restricted Keplerian motion about the asteroid, is
given. In Section 3, the expressions for the deflection are expressed
in terms of universal variables for the time of flight as these
simplify the optimization analysis that requires a calculation of the
time of flight on the restricted Keplerian orbit segment. Next in Section 4 a
new set of variables are introduced to simplify the numerical analysis
of the problem. The mass optimization problem is then 
formulated in Section 5, where the method of solution using dynamic
programming (see e.g. \cite{kirk,lewis-book}) is also outlined. Lastly, in Section 6,  a concrete example 
of the method is given using a hypothetical deflection problem of asteroid 2007 VK184.

\section{Background}

In this section a short introduction is given to the "dynamic" (i.e. not stationary with respect to the asteroid) gravity
tractor that is the subject of the mass optimization study
presented in the following sections. Details on the gravity tractor,
which has been shown to be generally more efficient than a stationary
gravity tractor, 
can be found in \cite{ketema}.

\begin{figure}[htbp]
  \begin{center}
    \unitlength=.5in
    \begin{picture}(4,5)
     \put(2.4,2.3){$\theta$}
     \put(-1.6,-1){\includegraphics[scale=0.55]{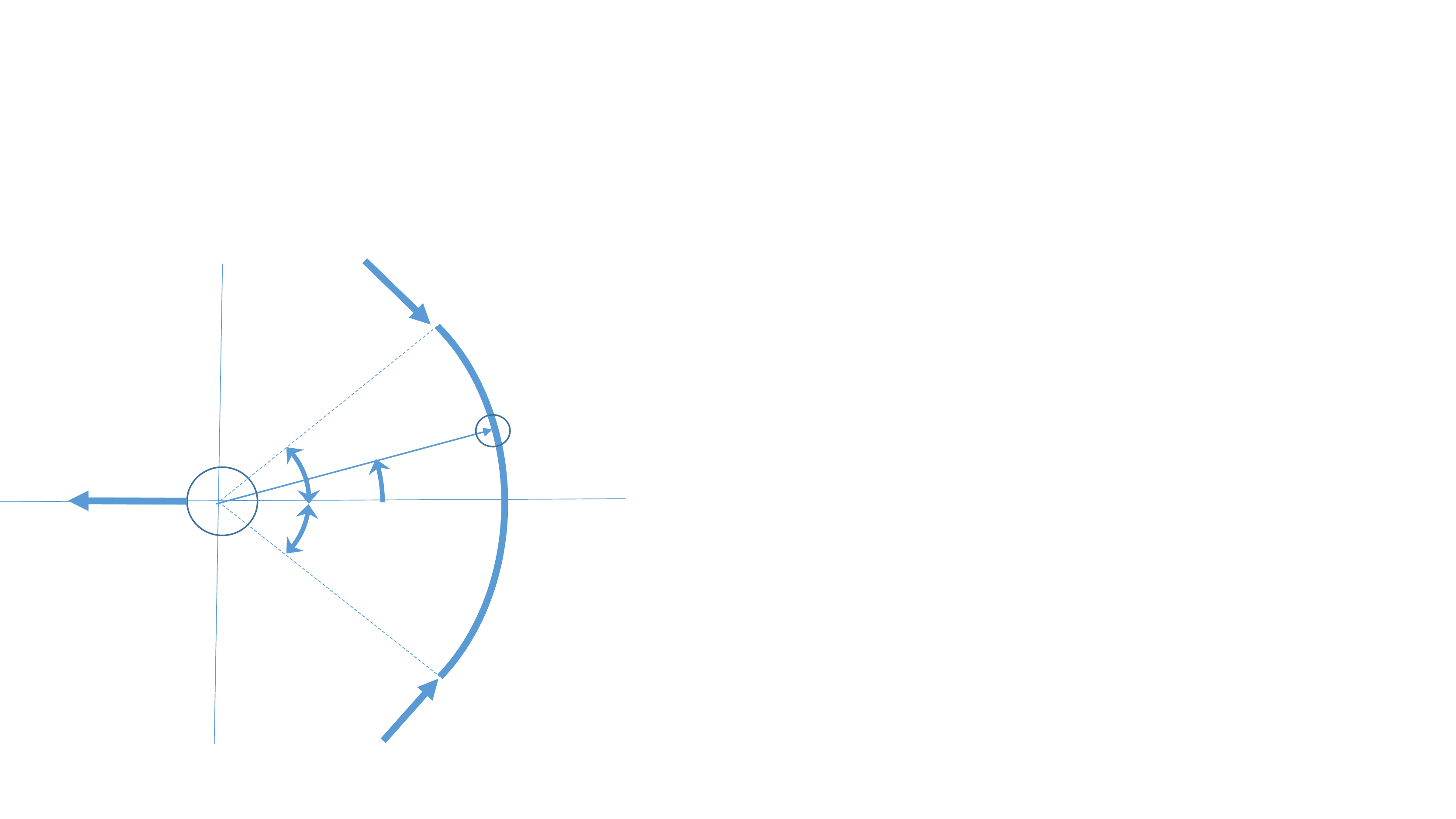}}
    \put(1.9,3.5){$r_{bi}$}
    \put(1.5,2.6){$\theta_{bi}$}
     \put(1.5,1.7){$\theta_{bi}$}
     \put(2.7,2.8){$r$}
     \put(-.5,2.6){asteroid}
     \put(-.8,1.8){$v_a$}
     \put(3.6,3.){spacecraft}
     \put(2.1,4.7){$\Delta v_{i+1}$}
     \put(1.9,.1){$\Delta v_{i}$}
     \put(3,.2){$A$}
     \put(2.9,4.1){$B$}
\put(3.6,1.9){$C$} 
\put(3.41,2.13){{$\bullet$}} 
     \put(4.9,2.1){$x$}
     \put(.4,4.7){$y$}

    \end{picture}
  \end{center}
  \caption{The asteroid-spacecraft system.}
\label{system}
\end{figure}
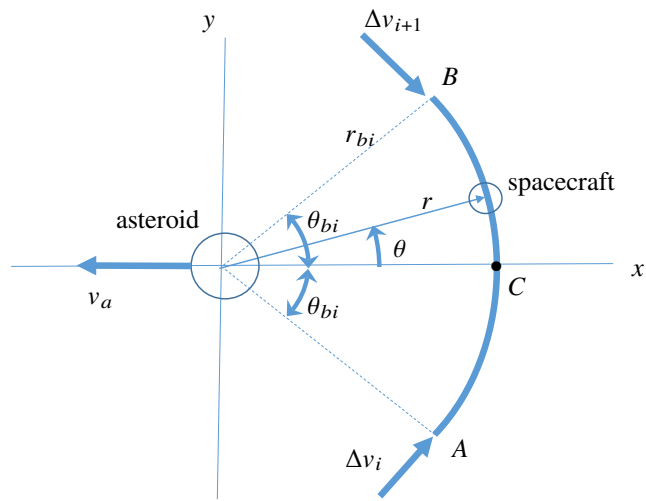

The dynamic gravity tractor consists of a spacecraft that moves back and forth
along a segment of a Keplerian orbit about the asteroid as shown
schematically in Figure \ref{system}. Assuming the orbit segment to be
symmetric about the $x$ axis, the average gravitational force on the
asteroid from the spacecraft over
one traversal of the orbit segment will be in the direction of the $x$
axis. This is the force that is used to effect a perturbation of
the asteroid's 
orbit about the Sun in order to avert a potential collision with
Earth. The intersection between the orbit segment and the $x$ axis, $C$, is an apsis
of the orbit segment: periapsis in the case of a parabolic or hyperbolic orbit, and
periapsis or apoapsis in the case of an elliptic orbit.

In order to change its direction of motion on the orbit segment, i.e.
at points $A$ and $B$ in Figure \ref{system}, the spacecraft
fires its thruster, thereby effecting  
an impulsive velocity change denoted by $\Delta v_{1i}$ and $\Delta v_{2i}$
at points $A$ and $B$, respectively. (The index $i$ corresponds to the number
of times the spacecraft has traversed the orbit segment, and reflects the fact
that the properties of the orbit segment may be different for each time.) Thus, due to the fuel expenditure
necessary for the impulsive velocity changes, the mass of the
spacecraft decreases on successive orbit segments.

The successive times at which the spacecraft is at the ends
of the orbit segment may be denoted by $t_i: i=1,2,3 ...$, such that
the start time for the asteroid deflection mission is $t_s=t_1$ and
the projected time of Earth-asteroid encounter is $t_e = t_N$.
The mass of the spacecraft between the times $t_i$ and $t_{i+1}$ will be denoted by $(m_c)_{i}$.

The asteroid is generally on an elliptic orbit about
the Sun. Therefore, its velocity is a function of time. However,
during the short period of time between $t_i$ and $t_{i+1}$ the
velocity may be assumed constant. This velocity will be
denoted by $v_{ai}$.

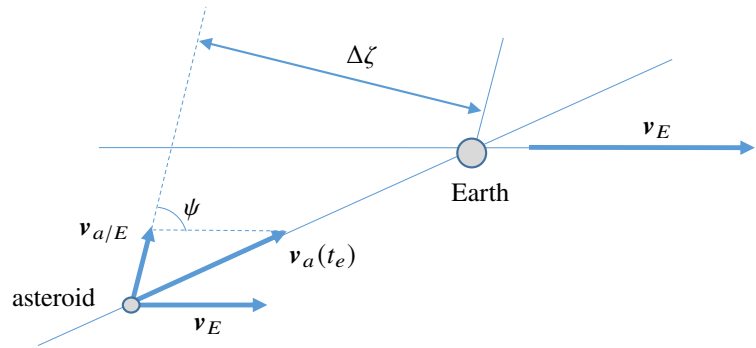
\begin{figure}[htbp]
  \begin{center}
    \unitlength=.5in
    \begin{picture}(4,3.8)
     \put(-2,-.85){\includegraphics[scale=0.8]{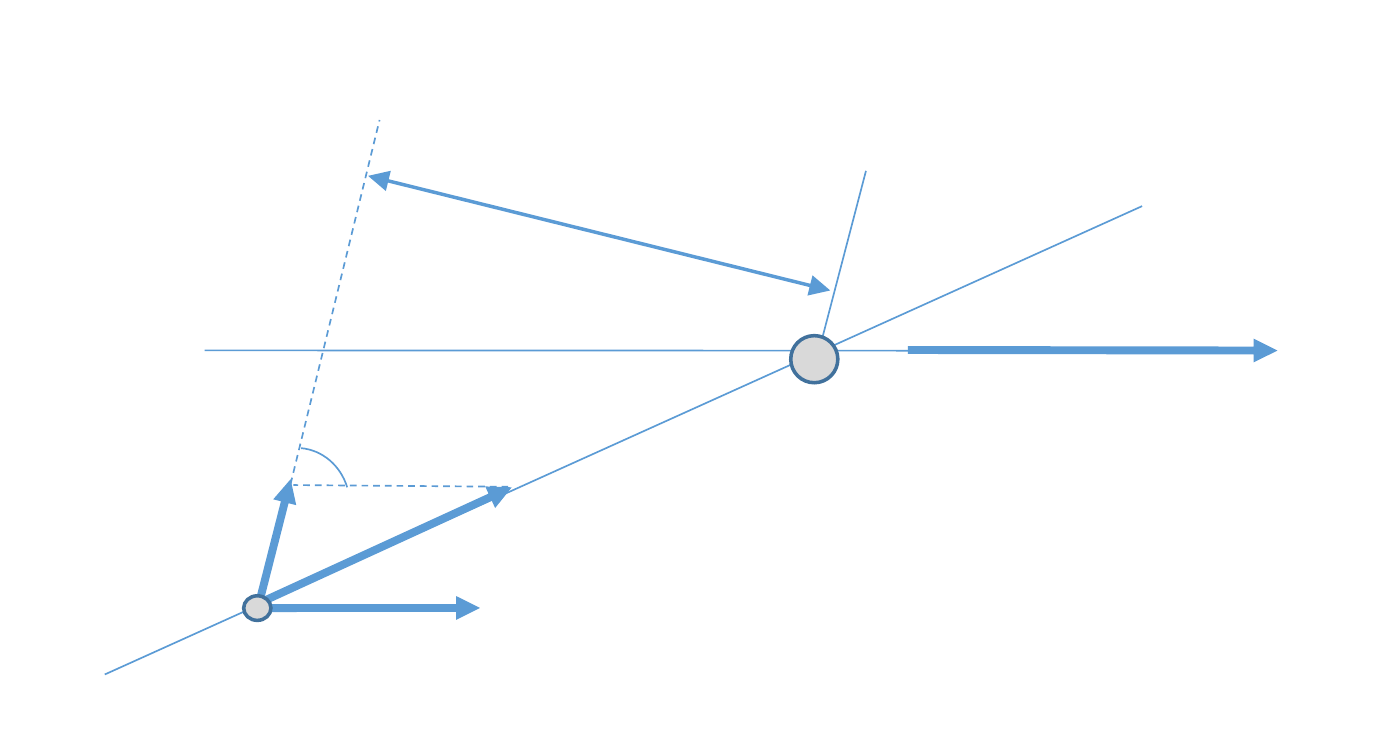}}
     \put(3,1.15){Earth}

    \put(5,1.85){$\bv_E$}
     \put(1.3,.5){$\bv_a(t_e)$}
     \put(-.9,.8){$\bv_{a/E}$}
     \put(-1.6,.05){asteroid}
     \put(1.9,2.6){$\Delta\zeta$}
     \put(.2,1.){$\psi$}
     \put(0.3,-.2){$\bv_E$}
    \end{picture}
  \end{center}
  \caption{Earth and the asteroid at the time of closest approach.}
\label{encounter}
\end{figure}

With the above definitions, for the $i$'th orbit segment, i.e. the one in the time interval
$[t_i,t_{i+1}]$ with
an angular momentum $h_i$ and bounding angle
$\theta_{bi}$, the contribution to the deflection of the
asteroid at the time of encounter $t_e$ may be written
as (see \cite{ketema})
\be
\Delta \zeta_i =-2G\kappa(m_c)_{i}v_{ai}(t_e-\bar{t}_i)\frac{\sin\theta_{bi}}{h_i}
\label{dz1}
\ee 
where $G$ is the universal gravitational constant and $\kappa$ is a constant defined through
\be
\kappa=\frac{3a_a}{\mu_S}v_a(t_e)\sin\psi,
\label{kappaeq}
\ee
with $a_a$ being the semimajor axis of the asteroid's heliocentric orbit,
$\mu_S$ is the gravitational parameter of the Sun, and $\psi$ the
angle between the heliocentric velocity of the asteroid and its
relative velocity with respect to Earth at the projected time of
encounter (see Figure \ref{encounter}).  Further,
$\bar{t}_i$ is defined as 
\be
\bar{t}_i=\frac{t_i+t_{i+1}}{2},
\ee
and can also be written as
\be
\bar{t}_i=t_i+\frac{\Delta t_i}{2},
\label{bt2}
\ee
where $\Delta t_i$ is the time of flight on the $i$'th orbit segment that can be calculated using Kepler's equation for the time of flight (see \cite{curtis3,bmw,hale}) and will be discussed further below.

The mass of the spacecraft $(m_c)_{i}$ will change every time
there is an impulsive velocity change $\Delta v_i$ in accordance with the relation
(see \cite{curtis2})
\be
(m_c)_{(i+1)}=(m_c)_{i}e^{-\beta\Delta v_i}
\label{dm}
\ee
where the parameter $\beta$ is defined through
\be
\beta=\frac{1}{I_{sp}g_0}
\ee
where $I_{sp}$ is the specific impulse of the spacecraft's thruster and
$g_0$ is the acceleration due to gravity at the surface of Earth \cite{hale2}.
The impulsive velocity change $\Delta v_i$ corresponds to a change in the
direction of the velocity at the ends of the orbit segment. Therefore
\be
\Delta v_i=2 v_{bi}
\label{vi}
\ee
where $v_{bi}$ is the velocity of the spacecraft at the end of the orbit segment.
Further, in the case of a small body such as an
asteroid, as will be seen in Section 6 (see also \cite{ketema}), $v_{bi}$ is generally small. This implies that one can make the assumption
\be
\beta\Delta v_i<<1
\label{obs}
\ee
and consequently (\ref{dm}) may be written in the form
\be
(m_c)_{(i+1)}=(m_c)_{i}(1-\beta\Delta v_i)
\label{dm2}
\ee

The total deflection caused at the time of encounter $t_e$ due to the
action of the gravity tractor from the initial time to a time
$t_k$ can be found as a sum of $\Delta \zeta_i$ which is given in (\ref{dz1}). Denoting this cumulative deflection by $\zeta_k$ it follows that
\be
\zeta_k=\sum_{i=1}^{k-1}\Delta\zeta_i\;\;\;\;k> 1
\label{zetak0}
\ee
where $\zeta_{1}=0$, or using (\ref{dz1})  
\be
\zeta_k=\sum_{i=1}^{k-1}-2G\kappa(m_c)_{i}v_{ak}(t_e-\bar{t}_k)\frac{\sin\theta_{bk}}{h_k}\;\;\;\;k> 1
\label{zetak}
\ee
It is also useful to note that
\be
\zeta_{k+1}=\zeta_k-2G\kappa(m_c)_{k}v_{ak}(t_e-\bar{t}_k)\frac{\sin\theta_{bk}}{h_k}
\label{zetarec}
\ee
thus giving a recursion formula for the deflection $\zeta_k$.

\subsection{Avoiding plume impingement}
For the gravitational force from the spacecraft on the asteroid to be
considered an external force that can have a net effect on the
asteroid's heliocentric motion (and on the asteroid-spacecraft
system's motion) it is necessary
for the exhausts from the asteroid's thruster to leave the asteroid's
sphere of influence without impinging on the asteroid, \cite{lu-love,ketema}.

\begin{figure}[htbp]
  \begin{center}
    \unitlength=.5in
    \begin{picture}(4,6.5)
\put(-2,-.5){\includegraphics[]{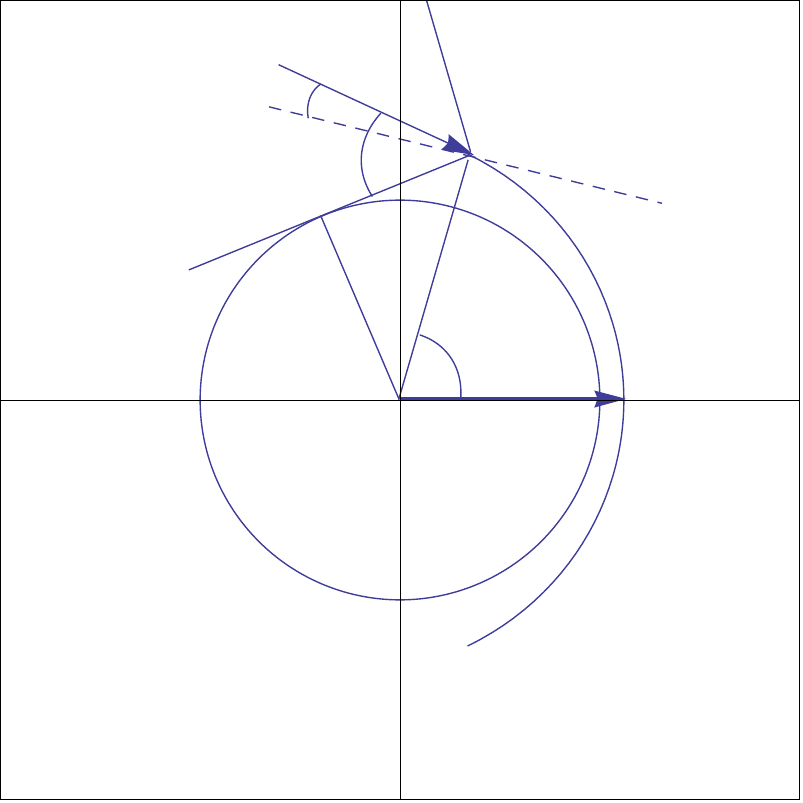}}
     \put(1.5,3.5){\small{$r(\theta_{bi})$}}
     \put(1.8,2.9){\small{$\theta_{bi}$}}
     \put(1.8,2.5){\small{$r_{ci}$}}
     \put(.15,5.15){\small{$\gamma_i$}}
     \put(.3,5.45){\small{$\Delta v_{i}$}}
     \put(.6,4.5){\small{$\varphi$}}
     \put(.55,3.5){\small{$r_a$}}
    \end{picture}
  \end{center}
  \caption{Effect of plume angle on minimum periapsis radius.}
\label{plumefig}
\end{figure}
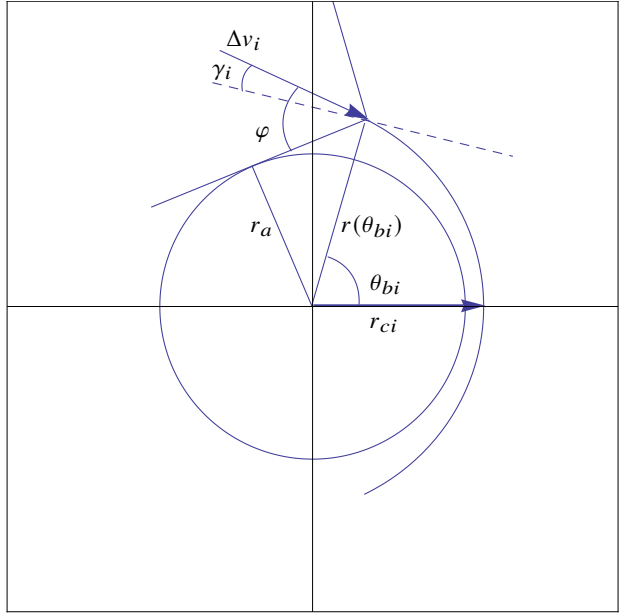

Figure \ref{plumefig} illustrates for the $i'$th orbit segment how the plume half-angle $\varphi$, the flight path angle $\gamma_i$, and the distance $r_{bi}$ determine whether or not there is plume impingement.
It follows from Figure(\ref{plumefig}) that the constraint that must be satisfied to avoid plume-impingement is
\be
r_{bi}\sin{(\frac{\pi}{2}-(\varphi-\gamma_i))}>r_a
\label{rpmt}
\ee
or
\be
r_{bi}\cos{(\varphi-\gamma_i)}>r_a
\label{rpm}
\ee
The flight path angle $\gamma$ satisfies (see \cite{curtis2})
\be
\tan\gamma=\frac{\dot{r}}{r\dot\theta}
\label{gamma0}
\ee 
Equivalently, noting that angular momentum $h$ may generally be written in the form (see \cite{curtis2})
\be
h=r^2\dot\theta
\label{ech}
\ee
(\ref{gamma0}) may be written as
\be
\tan\gamma=\frac{r\dot{r}}{h}
\label{gamma1}
\ee
and therefore for the $i'$th orbit segment
\be
\tan\gamma_i=\frac{r_{bi}\dot{r}_{bi}}{h_i}
\label{gamma}
\ee

\section{Universal variables}
In optimizing the initial gross mass of the spacecraft for a given required deflection of the asteroid within a given time interval, it will be necessary to
consider the time of flight on each orbit segment, i.e. from one end of the orbit segment to the
other. (This will be described further in the next section.) Because one needs to consider elliptic, hyperbolic, or
parabolic orbit segments in the general case, the time of flight
calculation is simplified 
if one uses universal variables \cite{bmw, curtis3}. This allows
for the determination of the time of flight using the same expression
for all three types of orbit segments. In this section, the required expressions for
the time of flight are considered. In the development of these
expressions, the index $i$ for the parameters
pertaining to the $i$'th orbit segment will be suppressed for brevity,
and will be reinstated at the end of the section.

\subsection{The time of flight}
In terms of universal variables, the time of flight $t$ from a point
$\br_0$ on an orbit of semimajor axis $a$ to a point $\br$ on the same
orbit may be written as \cite{bmw, curtis3} 
\be
\sqrt\mu t
=\chi^3S(z)+\frac{\br_0\cdot\bv_0}{\sqrt\mu}\chi^2C(z)+r_0\chi(1-zS(z))
\label{tof}
\ee
where $\chi$ is the universal variable for the time of flight,  $z$
is an auxilliary variable defined by
\be
z=\frac{\chi^2}{a}
\label{zee}
\ee
and where $r_0$ is defined through
\be
r_0=|\br_0|.
\ee
The Stumpff functions $C(z)$ and $S(z)$ are defined by
\ben
C(z)=\frac{1-\cos\sqrt{z}}{z}
\label{cofz}\\
S(z)=\frac{\sqrt{z}-\sin\sqrt{z}}{(\sqrt{z})^3}
\label{sofz}
\een
The value of $r$ corresponding to a given value of $\chi$ may be
written as (see e.g. \cite{bmw})
\be
r=\chi^2C(z)+\frac{\br_0\cdot\bv_0}{\sqrt\mu}(1-zS(z))\chi+r_0(1-zC(z))
\label{arr}
\ee
To apply the above expressions to calculate  the time of flight $\Delta
t$ between the two ends of the orbit segment, it may be noted that the
time of flight is twice the time of flight from the apsis $C$ 
to either end of the
orbit segment (see Figure \ref{system} ). Thus, in this case where the apsis of the orbit segment 
corresponds to $\br_0$, it follows that
\be
 \br_0\cdot\bv_0=0
\label{r0v0}
\ee
in  (\ref{tof}) and (\ref{arr}). The
time of flight between the two ends of the orbit segment may therefore
be written as
\be
\Delta t=\frac{2}{\sqrt{\mu}}\left[\chi^3S(z)+r_0\chi(1-zS(z))\right]
\label{tof2}
\ee
Next, using (\ref{r0v0}) in(\ref{arr}), the distance from the center of mass of the asteroid to the end of the
orbit segment may be written as
\be
r=\chi^2C(z)+r_0(1-zC(z))
\label{arr2}
\ee
Also, for later use, using (\ref{zee}), (\ref{cofz}),
and (\ref{r0v0}) in (\ref{arr2}) yields the expression
\be
r=a+(r_0-a)\cos\sqrt{z}
\label{rr0}
\ee

\subsection{The deflection in terms of universal variables}
In the expressions for the time of flight using universal variables, the parameters $r_0$, $a$, and 
$\chi$ (or $z$) characterize the orbit segment. 
Thus, for consistency in the overall problem, the deflection and conditions for avoiding
plume-impingement derived in Section 2 can also be expressed in terms of universal variables. 

In particular, regarding the deflection, the goal is to obtain an expression for the term
$\sin\theta_{bk}/h_k$ in (\ref{zetak}) and (\ref{zetarec}).
To this end, continuing to suppress the index $i$ for the $i$'th orbit segment, note that the 
radial component of the velocity may be written as (see \cite{curtis2}) 
\be
\dot{r}=\pm\frac{\mu}{h}e\sin\theta
\label{dotr0}
\ee
where '+' is used if $C$ is periapsis and '$-$' is used if $C$ is apoapsis of an elliptic orbit.
The expression $\sin\theta_b/h$ in (\ref{dz1}) may therefore be
written as
\be
\frac{\sin\theta}{h}=\pm\frac{\dot{r}}{\mu e}
\label{stb}
\ee
Next, in terms of univeral variables, $\dot{r}$ may be written as (see \cite{bmw})
\be
\dot{r}=e\frac{\sqrt{a\mu}}{r}\cos\frac{\chi+c_0}{\sqrt{a}}
\label{dotr}
\ee
or
\be
\dot{r}=e\frac{\sqrt{a\mu}}{r}\left[\cos\frac{\chi}{\sqrt{a}}\cos\frac{c_0}{\sqrt{a}}-\sin\frac{\chi}{\sqrt{a}}\sin\frac{c_0}{\sqrt{a}}\right]
\label{ardot}
\ee
where the constant $c_0$ satisfies the relations
\ben
e\sin\frac{c_0}{\sqrt{a}}=\frac{r_0}{a}-1
\label{c1}\\
e\cos\frac{c_0}{\sqrt{a}}=\frac{\br_0\cdot\bv_0}{\sqrt{\mu a}}
\label{c2}
\een
Using (\ref{r0v0}) in (\ref{c2}) and substituting (\ref{c1}) and
(\ref{c2}) into (\ref{ardot}) now gives
\be
\dot{r}=-\frac{\sqrt{a\mu}}{r}\left(\frac{r_0}{a}-1\right)\sin\frac{\chi}{\sqrt{a}}
\label{rdotf}
\ee
Next, substituting (\ref{rdotf}) in (\ref{stb}) gives
\be
\frac{\sin\theta}{h}=\mp\frac{1}{\mu e}\frac{\sqrt{\mu a}}{r}\left(\frac{r_0}{a}-1\right)\sin\sqrt{z}
\label{sintheta1}
\ee
Lastly, noting that 
\be
\frac{r_0}{a}-1=\mp e
\ee
where the top sign corresponds to the apsis $C$ in Figure \ref{system} being periapsis and the bottom sign to apoapsis  (in the case of an elliptic orbit) (\ref{sintheta1}) evaluated at the end of the orbit segment (i.e. $\theta=\theta_b$ etc.) may be written as
\be
\frac{\sin\theta_b}{h}=\sqrt{\frac{a}{\mu}}\frac{\sin\sqrt{z_b}}{r_b}
\label{sintheta}
\ee
which can be used in (\ref{dz1}) to give (using parameter values for the $i$'th orbit segment)
\be
\Delta \zeta_i =-2G\kappa v_{ai}\sqrt{\frac{a_i}{\mu}}\frac{1}{r_{bi}}(m_c)_{i}(t_e-\bar{t}_i)\sin\sqrt{z_{bi}}
\label{dz4}
\ee
and using (\ref{zetak0})
\be
\zeta_k=\sum_{i=1}^{k-1}-2G\kappa v_{ai}\sqrt{\frac{a_i}{\mu}}\frac{1}{r_{bi}}(m_c)_{i}(t_e-\bar{t}_i)\sin\sqrt{z_{bi}}
\label{zeta_univ}
\ee

\subsection{The plume non-impingement constraint}

Next, the condition to avoid
plume impingement (\ref{rpm}) will be expressed in terms of  the orbit segment parameters
$r_{0i}$, $a_i$, and $\chi_i$ (or $z_i$). Continuing to suppress  the index $i$ for simplicity and
expanding the left hand side of (\ref{rpm}) one obtains
\be
r_b\left[\cos\varphi\cos\gamma_b+\sin\varphi\sin\gamma_b\right]>r_a
\ee
or
\be
r_b\cos\gamma_b\left[\cos\varphi+\sin\varphi\tan\gamma_b\right]>r_a
\label{const}
\ee
Noting that the angular momentum can be found as (see \cite{curtis2})
\be
h=rv\cos\gamma
\ee
the expression for $r_b\cos\gamma$ at the end of the orbit segment is obtained as
\be
r_b\cos\gamma_b=\frac{h}{v_b}
\label{cosgamma}
\ee
which can be used in (\ref{const}) along with (\ref{gamma1}) to yield
\be
\frac{h}{v_b}\left[\cos\varphi+\sin\varphi\frac{r_b\dot{r}_b}{h}\right]>r_a
\label{const2}
\ee
The angular momentum for the orbit segment may also be calculated at the apsis 
as 
\be
h=r_0v_0
\label{angmom}
\ee
and using this expression in  (\ref{const2}) now gives
\be
(r_0v_0\cos\varphi+\dot{r_b}r_b\sin\varphi)>r_av_b
\label{const3}
\ee
Using the energy equation (written at the apsis) (see e.g. \cite{curtis2})
\be
\frac{v_0^2}{2}-\frac{\mu}{r_0}=-\frac{\mu}{2a}
\label{energy}
\ee
it follows that
\be
v_0=\sqrt{\frac{2\mu}{r_0}-\frac{\mu}{a}}
\label{v0}
\ee
and similarly at the end points of the segment
\be
v_b=\sqrt{\frac{2\mu}{r_b}-\frac{\mu}{a}}
\label{vb}
\ee
The above expressions for $v_0$ and $v_b$ can now be used in (\ref{const3}), and making use of (\ref{rdotf}), the
plume non-impingement condition may be written as
\be
r_0\sqrt{\frac{2\mu}{r_0}-\frac{\mu}{a}}\cos\phi-\sin\phi\sqrt{\frac{\mu}{a}}(r_0-a)\sin\frac{\chi_b}{\sqrt{a}}>r_a\sqrt{\frac{2\mu}{r_b}-\frac{\mu}{a}}
\label{plume}
\ee

where, using (\ref{rr0}),  $r_b$ is given by
\be
r_b=a+(r_0-a)\cos\sqrt{z_b}
\label{arbee1}
\ee
To avoid problems with the numerical representation $a$ in the case of a parabolic orbit segment
(i.e. $a->\infty$), a new variable $\alpha$ may be defined through (see e.g. \cite{curtis3},\cite{bmw})
\be
\alpha=\frac{1}{a}
\label{a2alpha}
\ee
Lastly, using (\ref{plume})-(\ref{a2alpha}) and defining the function
\ben
&&\Pi(r_{0},\alpha,\chi_b)=\nonumber\\
&&r_a\sqrt{\frac{2\mu}{r_b(r_0,\alpha,\chi)}-\mu\alpha}-r_0\sqrt{\frac{2\mu}{r_0}-\mu\alpha}\cos\phi+\sin\phi\sqrt{\mu\alpha}(r_0-\frac{1}{\alpha})\sin\chi_b\alpha^{\frac{1}{2}}
\label{Pi}
\een
where 

\be
r_b(r_0,\alpha,\chi)=\alpha^{-1}+(r_0-\alpha^{-1})\cos\chi\sqrt{\alpha}
\label{arbee2}
\ee
(\ref{plume}) may be written in the form
\be
\Pi(r_0,\alpha,\chi_{bi})<0
\label{noplume2}
\ee

\section{A change of variables}
As indicated earlier, the goal of the optimization analysis in the next section
will be to choose the values of the orbital parameters $r_{0k}$, $\alpha_k$, $\chi_{bk}$, $k=1\dots N-1$  
such that a desired amount of deflection of the asteroid is achieved
within a given 
amount of time and for the smallest possible initial mass of the
spacecraft. This problem, that will be defined more precisely in the
next section, takes on the form of a discrete-time nonlinear
optimal control problem where the orbital parameters $r_{0k}$, $\alpha_k$, and $\chi_{bk}$ constitute
the $3(N-1)$ control variables to be determined. Now, given
that the duration of the mission is in the order of years, while the time of flight on the orbit segment is in the order of an hour, $N$ will be large. For example, a 10-year-mission
with the time of flight on an orbit segment $\Delta t=30$ minutes gives $N=175,200$. This is significant because it would generally make a numerical approach to solving the problem
computationally intensive and slow.

The goal of this section is therefore to show that a different indexing of the variables can be
used that leads to a problem that is numerically more tractable.

\begin{figure}[htbp]
  \begin{center}
    \unitlength=.5in
    \begin{picture}(4,4)
     \put(-3,-.6){\includegraphics[scale=0.75]{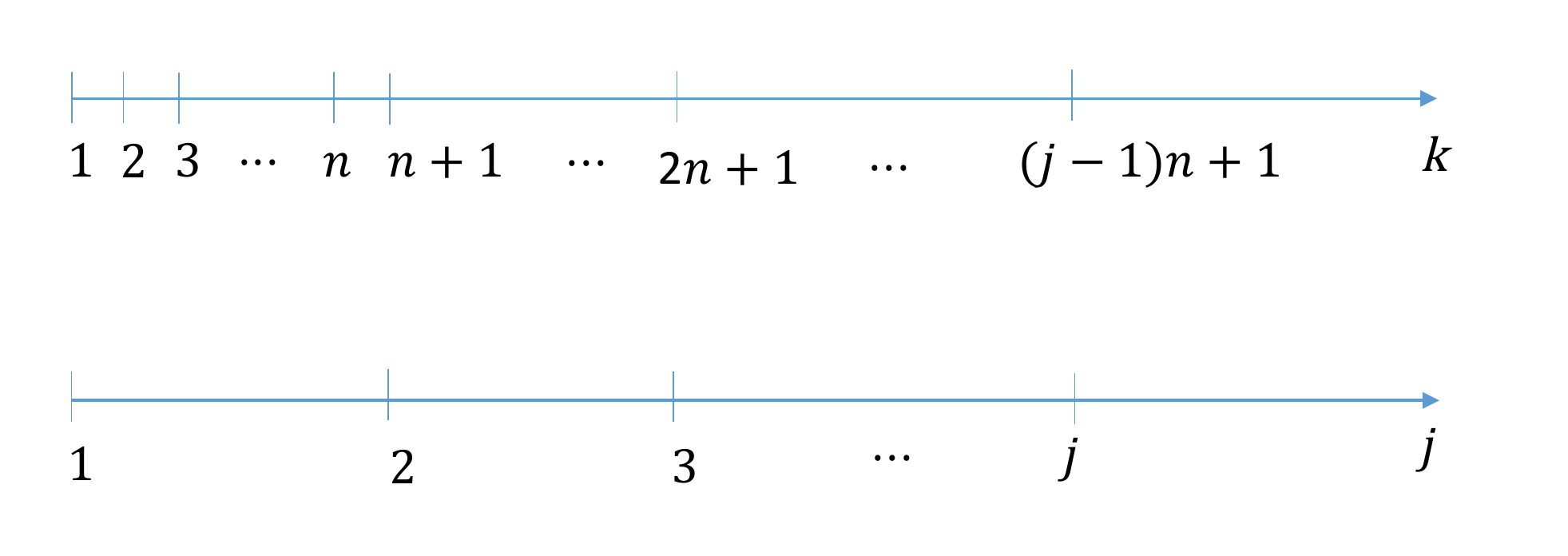}}
     \put(-2.5,1){$\tau_1$}
     \put(-2.5,1.4){$\eta_1$}
     \put(-2.5,3.2){$t_1$}
     \put(-2.5,3.6){$\zeta_1$}
     \put(-.2,1){$\tau_2$}
     \put(-.2,1.4){$\eta_2$}
     \put(-.2,3.2){$t_{n+1}$}
     \put(-.2,3.6){$\zeta_{n+1}$}
     \put(2,3.6){$\zeta_{2n+1}$}
     \put(2,3.2){$t_{2n+1}$}
     \put(2,1){$\tau_3$}
     \put(2,1.4){$\eta_3$}
     \put(5,3.6){$\zeta_{(j-1)n+1}$}
     \put(5,3.2){$t_{(j-1)n+1}$}
     \put(5,1){$\tau_j$}
     \put(5,1.4){$\eta_j$}
    \end{picture}
  \end{center}
  \caption{Relation between the indices $i$ and $j$.}
\label{ij}
\end{figure}
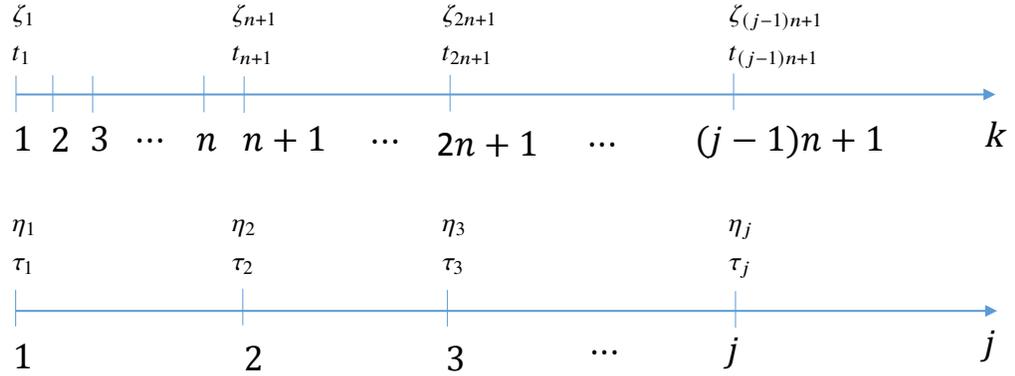

Thus note that even though $k$ was initially a natural choice as the index because it
corresponds to the number of times the spacecraft has traversed the
orbit segment, the variables $\zeta_k$, $\bar{t}_k$, and $m_k$ vary 
slowly with $k$, i.e. over a time scale in the order of years. In other words, it takes many traversions of the spacecraft's orbit segment before any
significant effect is observed in the variables $\zeta_k$, $\bar{t}_k$, and $m_k$. At the same time, the optimal values of the control variables at a given time
will generally depend on the values of $\zeta_k$, $t_k$, and $m_k$.
This means that the control variables will also vary slowly with $k$ and do not need to be updated after every traversion of the orbit segment.  Instead, they may be assumed to remain
constant for some chosen increment of $k$, say $n$. Hence, a new ``slowly varying'' index $j$ can be defined such that  $j$ increases by one for an increase of $k$ by $n$ (see Figure \ref{ij}) or
\be
j=\mbox{int}(\frac{k-1}{n})+1
\label{jaydef}
\ee
where $\mbox{int}(\cdot)$ is the integer part of the argument, and both $j$ and $k$ start at 1.  For ease of reference,
new state variables $\eta_j$, $\tau_j$, and $m_j$ corresponding to index $j$ may be defined. These variables are related to
the original state variables through
\ben
\eta_j=\zeta_{(j-1)n+1}
\label{zjay}\\
\tau_j={t}_{(j-1)n+1}\\
m_j=(m_c)_{(j-1)n+1}
\label{emjay}
\een
Figure \ref{ij} illustrates the definition of $\eta_j$ and $\tau_j$; the definition for $m_j$ is analogous.
Similarly, the control variables corresponding to the index $j$ are defined as $\tilde{r}_{0j}$, $\tilde{\alpha}_j$, and $\tilde{\chi}_j$ as follows:
\ben
\tilde{r}_{0j}=r_{(j-1)n+1}\\
\tilde{\alpha}_j=\alpha_{(j-1)n+1}\\
\tilde\chi_{bj}=\chi_{b[(j-1)n+1]}
\een
Conversely, for $k\in[(j-1)n+1,jn]$
\ben
r_{0k}=\tilde{r}_{0j}
\label{r0j}\\
\alpha_k=\tilde{\alpha}_{j}
\label{dakj}\\
\chi_{bk}=\tilde\chi_{bj}
\label{dxkj}
\een

\subsection{Recursion formulas}

As with the original state variables, state propagation equations for the variables $\eta_j$, $\tau_j$, and $m_j$ will need to be obtained. This can be done by evaluating the corresponding changes in the variables $\zeta_k$, $t_k$, and $(m_c)_k$ as will be shown in this section.

\subsubsection{The time variable:}
The recursion equation for the time variable takes
\be
\tau_{j+1}=\tau_j+n\Delta\tilde{t}_j
\label{tauj}
\ee
where $\Delta\tilde{t}_j$ denotes the time of flight on each orbit segment between the times of $j$ and $j+1$, or equivalently between $k=(n-1)j+1$ and $k=nj$. This time of flight is given by (\ref{tof2}) which can be written in terms of the variables $\tilde{r}_{0j}$, $\tilde{\alpha}_j$, and $\tilde{\chi}_j$ by using (\ref{zee}) and (\ref{a2alpha})  to yield
\be
\Delta {\tilde{t}}_j=\frac{2}{\sqrt{\mu}}\left[\tilde{\chi}_{bj}^3S(\tilde\chi_{bj}^2\tilde\alpha_j)+\tilde{r}_{0j}\tilde{\chi}_{bj}(1-\tilde{\chi}_{bj}^2\tilde{\alpha}_jS(\tilde{\chi}_{bj}^2\tilde{\alpha}_j)\right]
\label{tof3}
\ee

\subsubsection{The spacecraft mass:}
To determine the change in the mass of the spacecraft between time $j$ and $j+1$ we note that the thrusters will have been fired $n$ times, i.e. (\ref{dm2}) is used $n$ times with a constant value of $\Delta v_i$ that corresponds to the parameters of the orbit segment. Denoting this constant value by $\Delta \tilde v_j$ for this interval gives
\be
m_{j+1}=m_j(1-\beta\Delta \tilde v_j)^n
\label{mass1}
\ee
or, in keeping with the linearization used in obtaining (\ref{dm2})
\be
m_{j+1}=m_j(1-n\beta\Delta \tilde v_j)
\label{mass}
\ee

\subsubsection{The deflection}
Using (\ref{zjay}) and  (\ref{zeta_univ}) it follows that
\be
\eta_j=\sum_{i=1}^{(j-1)n}-2Gv_{ai}\sqrt{\frac{a_i}{\mu}}\frac{1}{r_{bi}}(m_c)_i(t_e-\bar{t}_i)\sin\sqrt{z_{bi}}
\label{etaj})
\ee
which implies that
\be
\eta_{j+1}=-\sum_{i=1}^{nj}2Gv_{ai}\sqrt{\frac{a_i}{\mu}}\frac{1}{r_{bi}}(m_c)_i(t_e-\bar{t}_i)\sin\sqrt{z_{bi}}
\label{zk0}
\ee
Using (\ref{zk0}) together with (\ref{etaj}) now gives
\be
\eta_{j+1}=\eta_j-\sum_{i=n(j-1)+1}^{nj}-2Gv_{ai}\sqrt{\frac{a_i}{\mu}}\frac{1}{r_{bi}}(m_c)_i(t_e-\bar{t}_i)\sin\sqrt{z_{bi}}
\label{zrec1}
\ee
and, noting that the orbital parameters of the orbit segment have constant values within the summation interval
given by (\ref{r0j})-(\ref{dxkj}), it follows that
\be 
\eta_{j+1}=\eta_j-2G\tilde{v}_{aj}\sqrt{\frac{\tilde{a}_j}{\mu}}\frac{1}{\tilde{r}_{bj}}\sin\sqrt{\tilde{z}_{bj}}\;\;\sigma_j
\label{zk2}
\ee
where
\be
\sigma_j=\sum_{i=n(j-1)+1}^{nj}(m_c)_{i}(t_e-\bar{t}_i)
\label{sigma}
\ee
has been defined for brevity.
In order to evaluate the sum in (\ref{sigma}), explicit expressions for the dependence of $(m_c)_i$ and $\bar{t}_i$ on $i$ are needed. Hence, it is first noted that the times of the impulsive thrusts $t_i$ within the summation interval [n(j-1)+1,nj] may be written as
\be
t_i=\tau_j+(i-n(j-1)-1)\Delta\tilde{t}_j, \;\;\;\; i\in[(j-1)n+1,nj]
\label{bt3}
\ee
and that the time of flight $\Delta t_i$ for each orbit segment within the summation interval is 
\be
\Delta t_i=\Delta\tilde t_j
\label{dtkj}
\ee
Further, using (\ref{bt2}) it follows that
\be
\bar{t}_i=\tau_j+(i-n(j-1)-1)\Delta \tilde{t}_{j}+\frac{\Delta\tilde{t}_j}{2};\;\;\;\;i\in[(j-1)n+1,jn]
\label{tandm1}
\ee
Similarly for the mass, using (\ref{dm2})
\be
(m_c)_{i}=(m_c)_{n(j-1)+1}(1-(i-n(j-1)-1)\beta\tilde\Delta v_j);\;\;\;i\in[(j-1)n+1,jn]
\label{m}
\ee
or using (\ref{emjay})
\be
(m_c)_{i}=m_{j}(1-(i-n(j-1)-1)\beta\Delta \tilde v_j);\;\;\;i\in[(j-1)n+1,jn]
\label{tandm2}
\ee
Introducing a new index $q$ through
\be
q=i-n(j-1)-1
\ee
and using (\ref{tandm1}) and (\ref{tandm2}) allows (\ref{sigma}) to be written in the form
\be
\sigma_j=\sum_{q=0}^{n-1}m_{j}(1-q\beta\Delta \tilde v_j)(t_e-\tau_j-q\Delta \tilde{t}_{j}-\frac{\Delta\tilde{t}_j}{2})
\ee
or,
after rearranging terms,
\be
\sigma_j=(t_e-\tau_j-\frac{\Delta\tilde{t}_j}{2})m_{j}\sum_{q=0}^{n-1}(1-q\beta\Delta
\tilde v_j)-m_j\Delta\tilde t_j\sum_{q=0}^{n-1}q(1-q\beta\Delta \tilde v_j)
\label{sum1}
\ee
Finally, evaluating the sums gives
\ben
\sigma_j=m_{j}(t_e-\tau_{j}-\frac{\Delta\tilde t_j}{2})\left(n-
\frac{n(n-1)}{2}\beta\Delta \tilde v_j\right)+
m_{j}\Delta \tilde
t_j\left(\frac{n(n-1)(2n-1)}{6}\beta\Delta \tilde v_j-\frac{n(n-1)}{2}\right)
\label{sum2}
\een

As has been discussed earlier, $t_e$ which is a time in the order of years
is much larger than $\Delta\tilde t_j$ which is in the order of
minutes. It is therefore possible to neglect the term
$\frac{\Delta\tilde t_j}{2}$ in the expression
$(t_e-\tau_j-\frac{\Delta\tilde t_j}{2})$ in (\ref{sum2}). Next, substituting (\ref{sum2}) into (\ref{zk2}) and using (\ref{zee}) and (\ref{a2alpha}) yields
\ben
\eta_{j+1}=\eta_j-2G\tilde{v}_{aj}\sqrt{\frac{\tilde{a}_j}{\mu}}\frac{1}{\tilde{r}_{bj}}\sin\sqrt{\tilde{z}_{bj}}
\times\nonumber\\
\left[m_{j}(t_e-\tau_{j})\left(n-
\frac{(n-1)}{2}\beta\Delta \tilde v_j\right)+
m_{j}\Delta \tilde
t_j\left(\frac{n}{6}(n-1)(2n-1)\beta\Delta \tilde v_j-\frac{n(n-1)}{2}\right)\right]
\label{zkf}
\een
\subsection{The non-impingement constraint for the plume}
Finally, the  plume non-impingement constraint
 (\ref{noplume2}) needs to be
specified for the $j$'th set of orbit segments, i.e. for
$i\in[(j-1)n+1,jn]$. However, this is simply a matter of replacing the
control variables $r_{0i}$, $\alpha_i$, and $\chi_i$ in
(\ref{noplume2}) with the new 
control variables defined in this section $\tilde{r}_{0j}$,
$\tilde\alpha_j$, and $\tilde\chi_j$ by using
(\ref{r0j})-(\ref{dxkj}). The plume non-impingement constraint is
therefore written  as
\be
\Pi(\tilde{r}_{0j},\tilde\alpha_j,\tilde\chi_{bj})<0
\label{noplume}
\ee

\section{Mass optimization}
The analysis in the rest of this paper is concerned with choosing the orbital parameters for 
the Keplerian orbit segment around the asteroid for each value of the discrete time $\tau_j$,
in order to effect a desired deflection at the projected time of Earth-encounter, while minimizing the required initial mass. 

As will be further detailed below, the following approach will be well-suited in conjunction with the use of the method of dynamic programming: For a range of values of initial mass of the spacecraft, the largest deflection that can be obtained by the time of the projected encounter is calculated. From this optimal mapping of initial mass to largest deflection, the smallest mass that results in the required deflection is picked. Thus in the optimization process itself, the goal is to minimize the deflection (with a direction corresponding to a negative value). However, only deflections that can be attained before the projected time of encounter are of value. Thus the cost function will include a large penalty for final times greater than $t_e$. The cost function may therefore be defined as
\be
J=\eta_N+Kf(t_N)
\label{pindex}
\ee
where $K$ is chosen so that $K>>\eta_N$ and
\be
f(t_N)=\left\{\begin{array}{cc}0&\;\;t_N\leq t_e\\1&\;\;t_N > t_e\end{array}\right.
\ee
\subsection{The control variables}
The parameters that characterize the orbit segment at each index $j$ are the distance $r_{0j}$ from the center of the asteroid to the apsis $C$,
the inverse of the semimajor axis $\tilde\alpha_j$, and the universal
variable at the end of the orbit segment $\tilde\chi_{bj}$ (or
equivalently $\tilde z_{bj}$). These are therefore the control variable that need to be determined at each index $j$. The $\tilde{} $ 's in these notations were
introduced in order to differentiate these quantities from the ones
that were updated after every traversion of the the orbit
segment. From this point on, however, the $\tilde{}$'s can be dropped for
simplicity; no confusion should arise as the original variables will
no longer be in use.

For use in what follows, the control vector may be defined as
\be
\bu_j=[r_{0j},\alpha_j,\chi_j]^T
\label{bu}
\ee

\subsection{The state variables}
The recursion equations for the state variables $m_j$, $\tau_j$, and $\eta_j$ were expressed in (\ref{mass}), (\ref{tauj}), and (\ref{zkf}), respectively.
Defining for brevity the functions
\be
Z(\eta_j,\tau_j,m_j;r_{0j},\alpha_j,\chi_{bj})=-2G\tilde{v}_{aj}\sqrt{\frac{\tilde{a}_j}{\mu}}\frac{1}{r(r_{0j},a_j,\chi_{bj})}\sin\sqrt{\tilde{z}_{bj}}
\ee
and
\ben
S(\eta_j,\tau_j,m_j;\Delta
v_j,\alpha_j,\chi_j)&=&\left[m_{j}(t_e-\tau_{j}-\frac{\Delta\tilde{t}_j}{2})\left(n-
\frac{(n-1)}{2}\beta\Delta v_j\right)+\right.\\
& &\left. m_{j}\Delta \tilde{t}_j\left(\frac{n}{6}(n-1)(2n-1)\beta\Delta v_j-\frac{n(n-1)}{2}\right)\right]
\een
where $\Delta v_j$ is calculated using (\ref{vi}) and (\ref{vb}),
(\ref{zkf}) may be written as
\be
\eta_{j+1}=\eta_j+Z(\eta_j,\tau_j,m_j;\Delta v_j,\alpha_j,\chi_j)S(\eta_j,\tau_j,m_j;\Delta
v_j,\alpha_j,\chi_j)
\ee
Next, using (\ref{tauj}) and (\ref{mass}) the three state equations can be
put in the form
\be
\left[\begin{array}{c}
\eta_{j+1}\\\tau_{j+1}\\m_{j+1}
\end{array}\right]=
\left[\begin{array}{c}
\eta_j+Z(\eta_j,\tau_j,m_j;\Delta v_j,\alpha_j,\chi_j)S(\eta_j,\tau_j,m_j;\Delta
v_j,\alpha_j,\chi_j)\\
\tau_{j}+n\Delta \tilde t_j\\
m_j(1-n\beta\Delta v_j)
\end{array}
\right]
\label{constraint1}
\ee
The $j'$th state vector $\bx_j$ may be defined through
\be
\bx_j=[\eta_j,\tau_j,m_j]^T
\label{bx}
\ee
and the state propagation equations may be written as
\be
\bx_{j+1}=\bff(\bx_j,\bu_j)
\label{state}
\ee
where, using (\ref{constraint1}), the vector function $\bff(\bx_j,\bu_j)$ is defined through
\be
\bff(\bx_j,\bu_j)=\left[\begin{array}{c}
\eta_j+Z(\eta_j,\tau_j,m_j;\Delta v_j,\alpha_j,\chi_j)S(\eta_j,\tau_j,m_j;\Delta
v_j,\alpha_j,\chi_j)\\
\tau_j+n\Delta \tilde{t}_j\\
m_j(1-n\beta\Delta v_j)
\end{array}
\right]
\label{bff}
\ee

\subsection{Constraints on the controls}
As discussed in Section 4, the controls $\tilde{r}_{0j}$, $a_j$, and $\chi_j$ are required to satisfy the
following constraints:
\subsubsection{No plume impingement}
This requirement has already been discussed in Section 3.4 and
stated in (\ref{noplume}). 

\subsubsection{No impact}
In addition to the avoidance of plume-impingement, it must be required
that the orbit segment itself does not 
intersect the asteroid. Thus if the apsis point
at the distance $r_0$ from the center of the asteroid is periapsis, then the no impact  condition is
\be
r_{0j}>r_a
\label{rcon1}
\ee
or

\be
r_a-r_{0j}<0
\label{rcon}
\ee
If on the other hand the orbit segment is elliptic and the apsis point is apoapsis, the points on the orbit segment closest to the asteroid are the end points at a distance of $r_b$ from the center of the asteroid. In this case the no impact condition becomes
\be
r_b>r_a
\ee
However, this condition is already satisfied due to the no plume-impingement condition given in (\ref{rpm}) and therefore does not need to be imposed separately.

\subsubsection{Sufficient time between impulsive velocity changes}
The assumption of impulsive
thrusts used in the analysis in this paper implies that the time of flight is much larger than the time
required in practice to impart the required $\Delta v$.
Hence, a lower-bound constraint on the time of flight is needed
and may be written as
\be
\Delta\tilde{t}_j>t_{min}
\label{tcon1}
\ee
or using (\ref{tof3})and defining the function
\ben
&&\Upsilon(r_{0j},\alpha_j,\chi_{bj})=\nonumber\\
&&t_{min}-\frac{2}{\sqrt{\mu}}\left[\tilde{\chi}_j^3S(\tilde\chi_j^2\tilde\alpha_j)+\tilde{r}_{0j}\tilde{\chi}_{bj}(1-\tilde{\chi}_{bj}^2\tilde{\alpha}_jS(\tilde{\chi}_{bj}^2\tilde{\alpha}_j)\right]
\label{upsilon}
\een
the constraint on the time of flight becomes
\be
\Upsilon(\Delta v_j,\alpha_j,\chi_{bj})<0
\label{tcon}
\ee

The three constraints on the controls that have been described may be expressed more compactly if one defines the vector
\be
G(r_{0j},\alpha_j,\chi_j)=\left[\begin{array}{c}r_a-r_{0j}\\
\Gamma(r_{0j},\alpha_j,\chi_{bj})\\
\Upsilon(r_{0j},\alpha_j,\chi_{bj})
\end{array}
\right]
\label{constraints2}
\ee
to give the overall constraint
\be
G(r_{0j},\alpha_j,\chi_{bj})<{\bf 0}
\label{Gl0}
\ee
\subsection{Dynamic programming}
Using (\ref{pindex}) it may be noted that the performance index is in the form \footnote{More generally the perfomance index would have the form $J=\phi(\bx_\nu)+\sum_{j=0}^{\nu-1}{\cal L}(\bx_j,\bu_j)$, i.e. would also depend on the history of the state and control variables (see e.g. \cite{kirk})}.
\be
J=\phi(\bx_\nu)
\label{jlag}
\ee
where $\nu$ is the final value of the index $j$.
This problem is straightforward to solve using the numerical approach
of dynamic programming \cite{kirk,lewis-book}.  Thus, for each index $j$ in
(\ref{jlag}) a
discretized set of values of the state variable $\bx_j$ and the
control variable $\bu_j$ are considered. The set of values of $\bx_j$ 
corresponding to a grid formed by discretizing the
deflection, mass, and time variables that may be denoted as
\ben
\bar{\eta}_j&=&\{\eta^1_j,\eta^2_j,\eta^3_j,\dots,\eta^{q_\eta}_j\}\\
\bar{\tau}_j&=&\{\tau^1_j,\tau^2_j,\tau^3_j,,\dots,\tau^{q_\tau}_j\}\\
\bar{m}_j&=&\{m^1_j,m^2_j,m^3_j,\dots,m^{q_m}_j\}
\een
where $q_\eta$, $q_\tau$, and $q_m$ denote the lengths of the arrays used for the variables
$\bar{\eta}$, $\bar{\tau}$, and $\bar{m}$, respectively.

Similarly, $\bu_j$ will have a set of of values on a grid formed by
discretized values of $\bar{r}_0$, $\bar{\alpha}$, and
$\bar{\chi}$ defined as
\ben
\bar{r}_{0j}&=&\{r^1_{0j},r^2_{0j},r^3_{0j},\dots,r^{q_{r_0}}_{0j}\}\\
\bar{\alpha}_j&=&\{\alpha^1_j,\alpha^2_j,+\alpha^3_j,\dots,\alpha^{q_\alpha}_j\}\\
\bar{\chi}_{bj}&=&\{\chi^1_{bj},\chi^2_{bj},\chi^3_{bj},\dots,\chi^{q_\chi}_{bj}\}
\een
where $q_{{r}_0}$, $q_\alpha$, and $q_\chi$ denote the lengths of the arrays
$\bar{r}_0$, $\bar{\alpha}$, and $\bar{\chi}$, respectively.

The optimization procedure is started by calculating and storing the contribution
from each of the final states $\bx_\nu$ to the performance
index on a trajectory that would end at that final state. This value may be denoted as
\be
J_\nu(\bx_\nu)=\phi(\bx_\nu)
\ee
Next, the index $j=\nu-1$ is considered for the discretized set of
states $\bx_{\nu-1}$. At each value of $\bx_{\nu-1}$ all possible
combinations of the controls are used to determine the
values of $\bx_\nu$ that they would result in, i.e., using (\ref{state})
\be
\bx_\nu=\bff(\bx_{\nu-1},\bu_{\nu-1})
\ee
The total cost
assosciated with such a trajectory is calculated as
\be
J_{\nu-1,\nu}(\bx_{\nu-1},\bu_{\nu-1})=J_\nu(\bx_\nu)
\ee
Of these values, the minimum is calculated for each $\bx_{\nu-1}$ and
is denoted as
\be
J^*_{\nu-1,\nu}(\bx_{\nu-1})=\min_{\bu_{\nu-1}}J_\nu(\bx_\nu)
\ee
In the process, the corresponding optimal value of $\bu_j(\bx_{\nu-1})$
is calculated and stored.

The procedure is now repeated for the states $\bx_{\nu-2}$ by
determining the controls that minimize the cost on trajectories from
each $\bx_{\nu-2}$ to corresponding final states $\bx_\nu$.and,
considering trajectories to the states $\bx_{\nu-1}$. The minimal cost
for each $\bx_{\nu-2}$ to go from $j=\nu-2$ to $j=\nu$  may be denoted as
\be
J^*_{\nu-2,\nu}(\bx_{\nu-2})=\min_{\bu_{\nu-2}} J_\nu(\bx_{\nu-1})
\ee
A repeat of the above procedure leads to a recursion formula in the
form
\be
J^*_{\nu-k,\nu}(\bx_{\nu-k})=\min_{\bu_{\nu-k}}J^*_{\nu-(k-1),\nu} (\bff(\bx_{\nu-k},\bu_{\nu-k}))
\label{recursion}
\ee

The end result of the algorithm is that for $k=\nu-1$ i.e. $\nu-k=1$,
the optimal cost $J^*_{1,\nu}(\bx_1)$ is obtained for each
value of $\bx_1$, i.e. the optimal cost on a trajectory that
starts at state $\bx_1$ and the associated controls $\bu^*_j$.  At this point, using the constraints that $\eta_1=0$
and $\tau_1=0$, only the various possible values of initial mass need
be considered. This defines a map from the initial values of gross
spacecraft mass to the corresponding deflection that minimizes $J$. 
\be
\eta_\nu=P(m_1)
\ee
The value of $m_1$ that corresponds to $\eta_\nu=d_r$ can now be calculated
by interpolation.
\subsection{Choosing the number of steps}
As described above, $\nu$ is the final value of the index $j$. Due to the varying values of $\Delta t_j$, however, $\nu$ cannot be simply derived
from the value of the final time $t_e$. Instead, $\nu$ can be chosen to be sufficiently large to allow $\tau_j$ to reach  $t_e$ for some value of $j=j_f$. Now, once $\tau_j \geq t_e$, there is no need to propagate the state variables any more and they can be assumed to stay constant at their values at $j=j_f$. 

\section{Example: Deflection of Asteroid VK184}
In order to exemplify the optimization procedure detailed above, and
to evaluate the merits of the approach, a 
concrete example is presented in this section concerning the
hypothetical deflection of an asteroid. The asteroid
considered is  2007 VK184, which is also considered in \cite{ketema}
and \cite{olympio}
and will therefore allow for a comparison of the results in these
works to those in the current paper.

Asteroid 2007 VK184 is projected to have a close encounter with Earth in June 2048,  with virtually no risk of collision. The mass and radius of the asteroid are $m_a=3.3\times10^9$ kg and $r_a=65$ m, respectively \cite{massradius}. The semimajor axis of the asteroid's orbit is $a_a=1.7262AU$ \cite{elements}.

For the parameters $\psi$, $v_a(t_e)$, and $a_a$ in (\ref{kappaeq}) it is shown in \cite{ketema} that
\ben
\psi&=&0.829 \mbox{ rad}\\
v_a(t_e)&=&35,500 \mbox{ m/s}
\label{parameters}
\een
which allows for the calculation of $\kappa$ as
\be
\kappa=1.528\times10^{-4} \mbox{ s/m}
\label{kappa_num}
\ee

For the sake of
specificity, it will be assumed in this section that
the encounter time $t_e=10$ years (i.e. the deflection
maneuvers would start in the year 2038) and that the smallest
allowable time of flight on the orbit segment is $\Delta t_{min}=1800$ s.

Denoting a one-dimensional array $\bar{p}$ of $N$ equally spaced values with first value $p_1$ and last value $p_N$ by
\be
\bar{p}={\cal A}(p_1,p_N,N)
\ee
the grid of values for $\bar\eta$, $\bar\tau$, and $\bar{m}$ in the dynamic programming algorithm are taken to be
\ben
\bar\eta&=&{\cal A}(0,-2.4\times 10^6 \mbox{m},20)\\
\bar\tau&=&{\cal A}(0,10 \mbox{yrs},20)\\
\bar{m}&=&{\cal A}(100,3600 \mbox{kg},20)
\een

\begin{figure}[htbp]
  \begin{center}
    \unitlength=.5in
    \begin{picture}(4,7)
     \put(-3.6,-3.8){\includegraphics[scale=0.65]{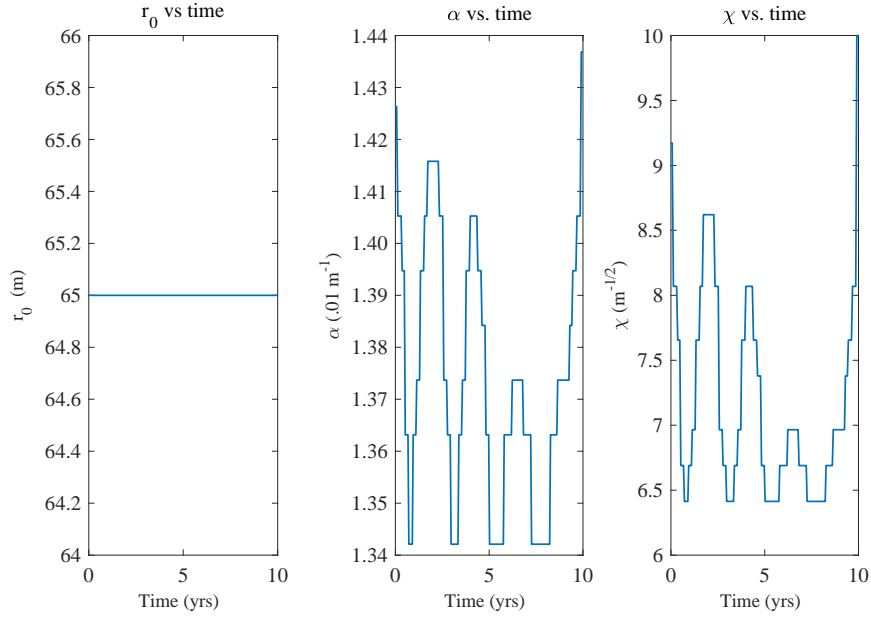}}
    \end{picture}
  \end{center}
  \caption{Time dependence of the control variables $r_0$, $\alpha$, and $\chi$ for the case of $t_e=10$ years.}
\label{cdep}
\end{figure}

Similarly, a grid of values can be chosen for the controls. However, iterative numerical studies show that the optimal value of $r_0$ is it's smallest allowable value, i.e. $r_e$, the radius of the asteroid. This may be explained by the fact that a small $r_0$ will generally increase the gravitational force. In the interest of computational time and higher resolution for the other control variables, $r_0$ is set to 65 m, the radius of the asteroid. For the other two control variables
\ben
\bar{\alpha}&=&{\cal A}(0.012/\mbox{m},0.015/\mbox{m},20)\\
\bar{\chi}&=&{\cal A}(6\;m^{1/2},10\;m^{1/2},30)
\een
The array sizes for $\bar{\alpha}$ and $\bar{\chi}$ were chosen iteratively in order to give the resolution required to adequately describe the variations of the variables as seen in Figure \ref{cdep}. 

Lastly, the parameter $n$ in (\ref{sum2}) is set to $n=700$. This choice is based on an assumption of $\Delta t_j \approx 1800$ s (the upper bound set) meaning that the total time corresponding to a unit increment of $j$ is roughly 1.26$\times 10^6$ s or about 14 days. This time interval is about 1.6\% of the period of the asteroid (2.34 years), making the assumption of a constant asteroid velocity during the time interval reasonable.

Applying dynamic programming as described above, the optimal trajectory starting from each admissible initial state of $\eta=0 \mbox{m}$, $\tau=0\mbox{s}$, and $m\in\bar{m}$
is calculated. Each value of the initial mass will result in a corresponding final value of the deflection $\eta_\nu$. The trajectory of interest is one that would result in a final value of $\eta=dr$. The initial mass corresponding to that trajectory can now be found by interpolation.

%

\begin{figure}[htbp]
  \begin{center}
    \unitlength=.5in
    \begin{picture}(4,7)
     \put(-3.6,-3.8){\includegraphics[scale=0.64]{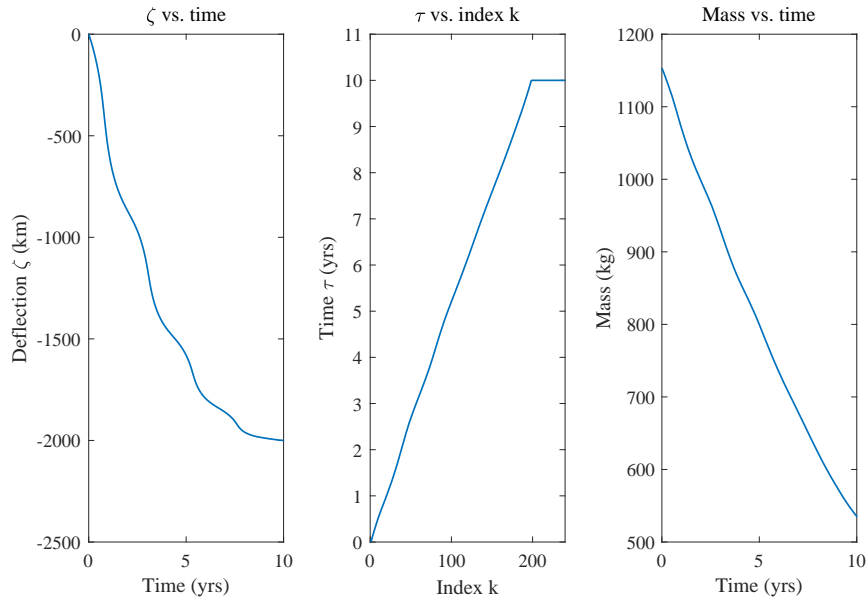}}
    \end{picture}
  \end{center}
  \caption{Example time dependence of the state variables $\eta$, $\tau$, and $m$ for the case of $t_e=10$ years.}
\label{tdep}
\end{figure}

The time dependence for the corresponding optimal solution of the
state variables is shown in Figures (\ref{tdep}). In particular it is seen from Figure (\ref{tdep})(c) that the optimal initial and final spacecraft masses are 1150 kg and 540 kg, respectively.

Similarly, Figure \ref{cdep} shows the optimal time dependence of the control variables $r_0$, $\alpha$, and $\chi$. The piecewise constant nature of the dependence for $\alpha$, and $\chi$ is a result of the discretization process for dynamic programming that has been described above.

\begin{figure}[htbp]
  \begin{center}
    \unitlength=.5in
    \begin{picture}(4,7.2)
     \put(-5.2,-5.8){\includegraphics[scale=0.85]{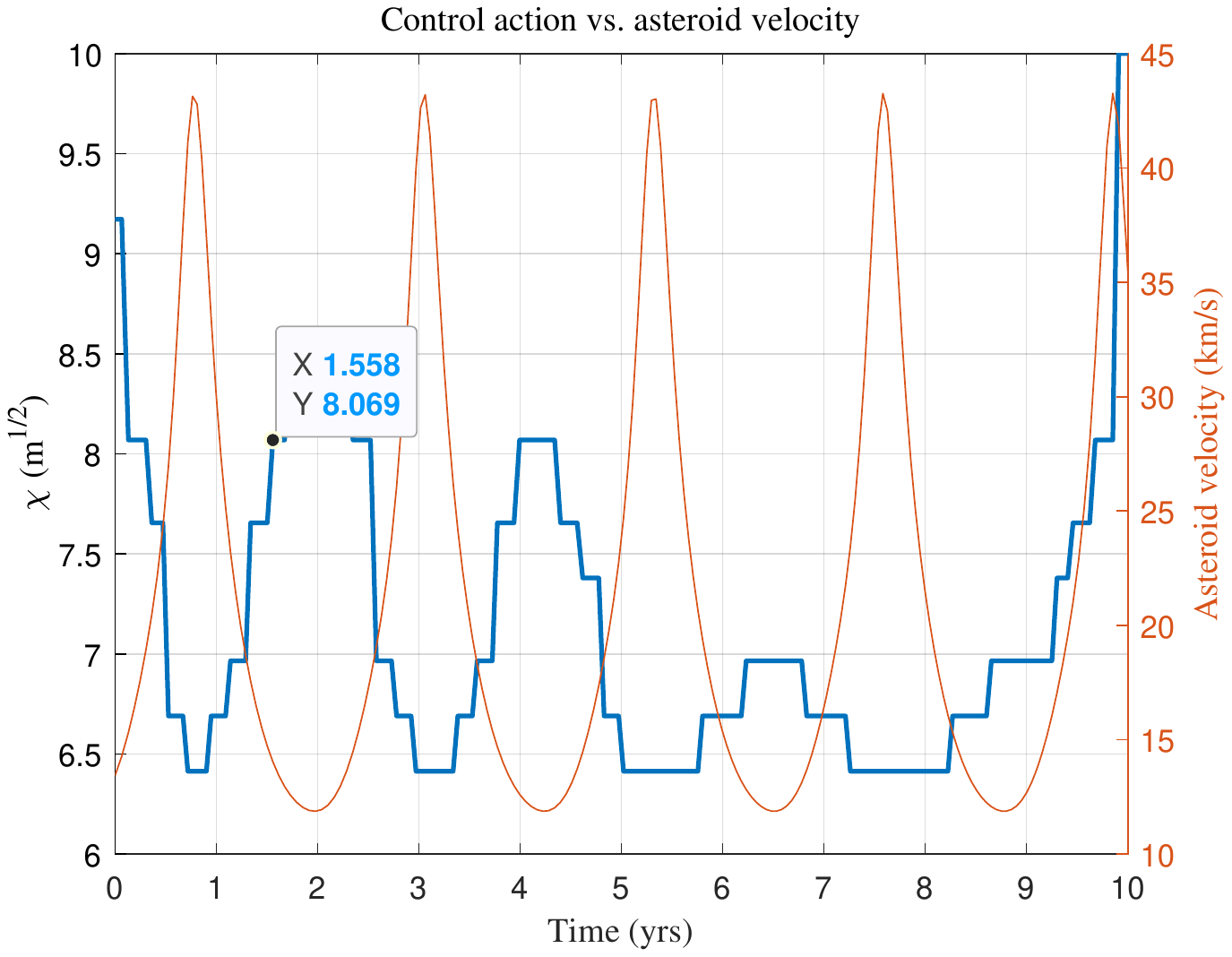}}
    \end{picture}
  \end{center}
  \caption{Time dependence of the control variable $r_0$, $\chi$ for the case of $t_e=10$ years and its relation to the asteroid velocity.}
\label{tcomp}
\end{figure}

The overall time dependence of the control variables $\alpha$, and $\chi$ may be understood in terms of the varying velocity of the asteroid. In Figure \ref{tcomp} the time-dependence of the asteroid velocity is shown along with the time-dependence of $\chi$ (note that $\alpha$ has the same qualitative time-dependence as $\chi$). It follows that $\chi$ (and therefore $\alpha$) have their smallest and largest values at the maximum and minimum asteroid velocities, respectively. A small $\alpha$ corresponds to a larger (higher energy) orbit where the velocities, and therefore changes in velocities, at the ends of the orbit segment are large. Correspondingly, a small $\chi$ means that the $\Delta v$'s are being applied at a higher rate, increasing the average force. It may be concluded that the overall control effort is larger when the asteroid velocity is larger, and vice versa. This is a result of the way in which a contribution to the deflection $\Delta \zeta_k$ depends on the asteroid velocity as expressed in (\ref{dz4}), making it more efficient to exert a larger force when the asteroid velocity is large.

\begin{figure}[htbp]
  \begin{center}
    \unitlength=.5in
    \begin{picture}(4,7.2)
     \put(-3.2,-.5){\includegraphics[scale=0.65]{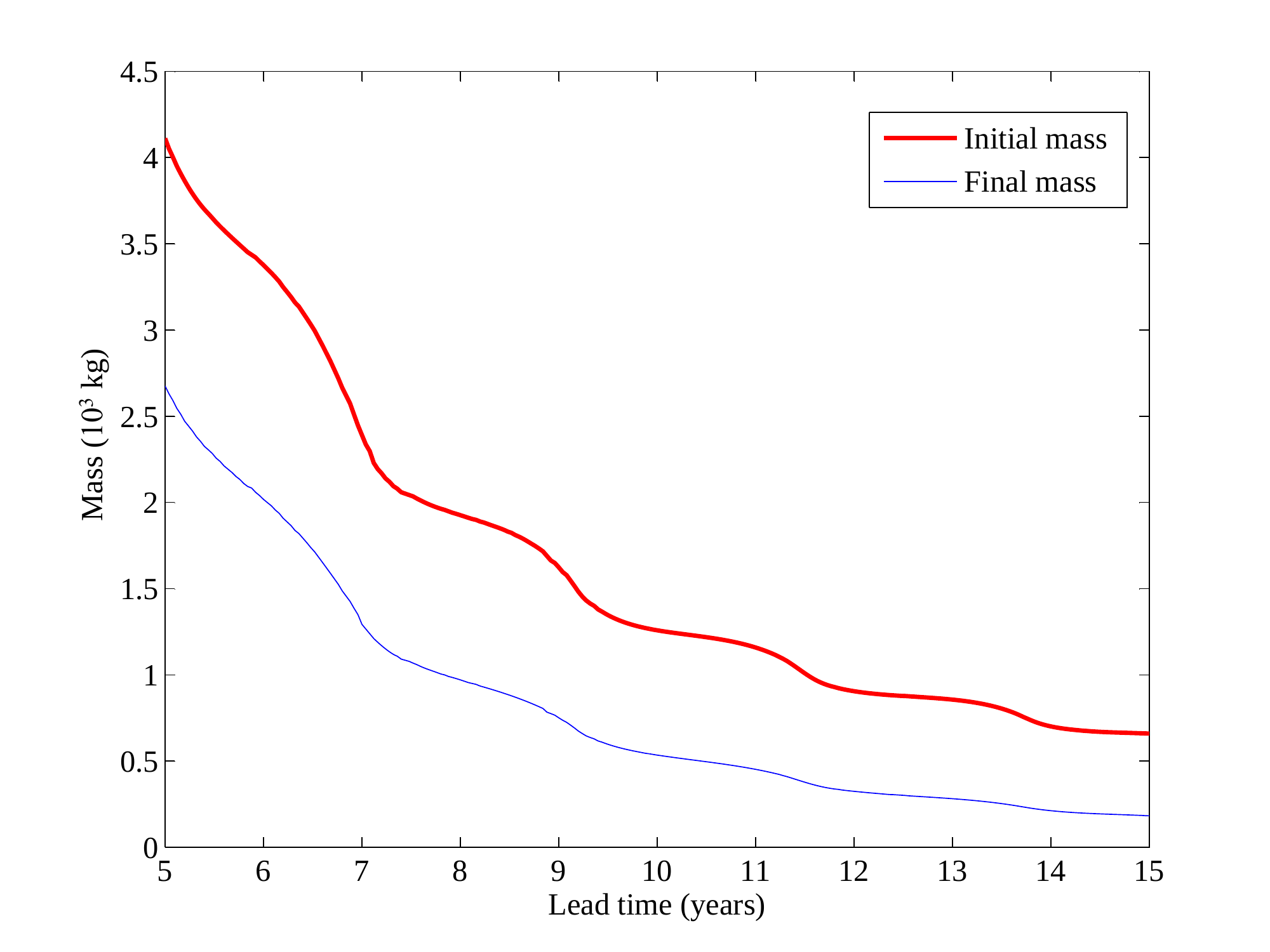}}
    \end{picture}
  \end{center}
  \caption{Optimal wet mass and dry mass for the gravity tractor spacecraft as a
    function of lead time at start of deflection action.}
\label{wetdry}
\end{figure}

To study the dependence of the required initial spacecraft mass on the
lead time for the start of the mission, the optimal trajectories can
be obtained for a range of values of $t_e$. Figure (\ref{wetdry})
shows this dependence of the optimal initial and final masses on the
lead time.

\begin{figure}[htbp]
  \begin{center}
    \unitlength=.5in
    \begin{picture}(4,7.5)
     \put(-3,-.5){\includegraphics[scale=0.65]{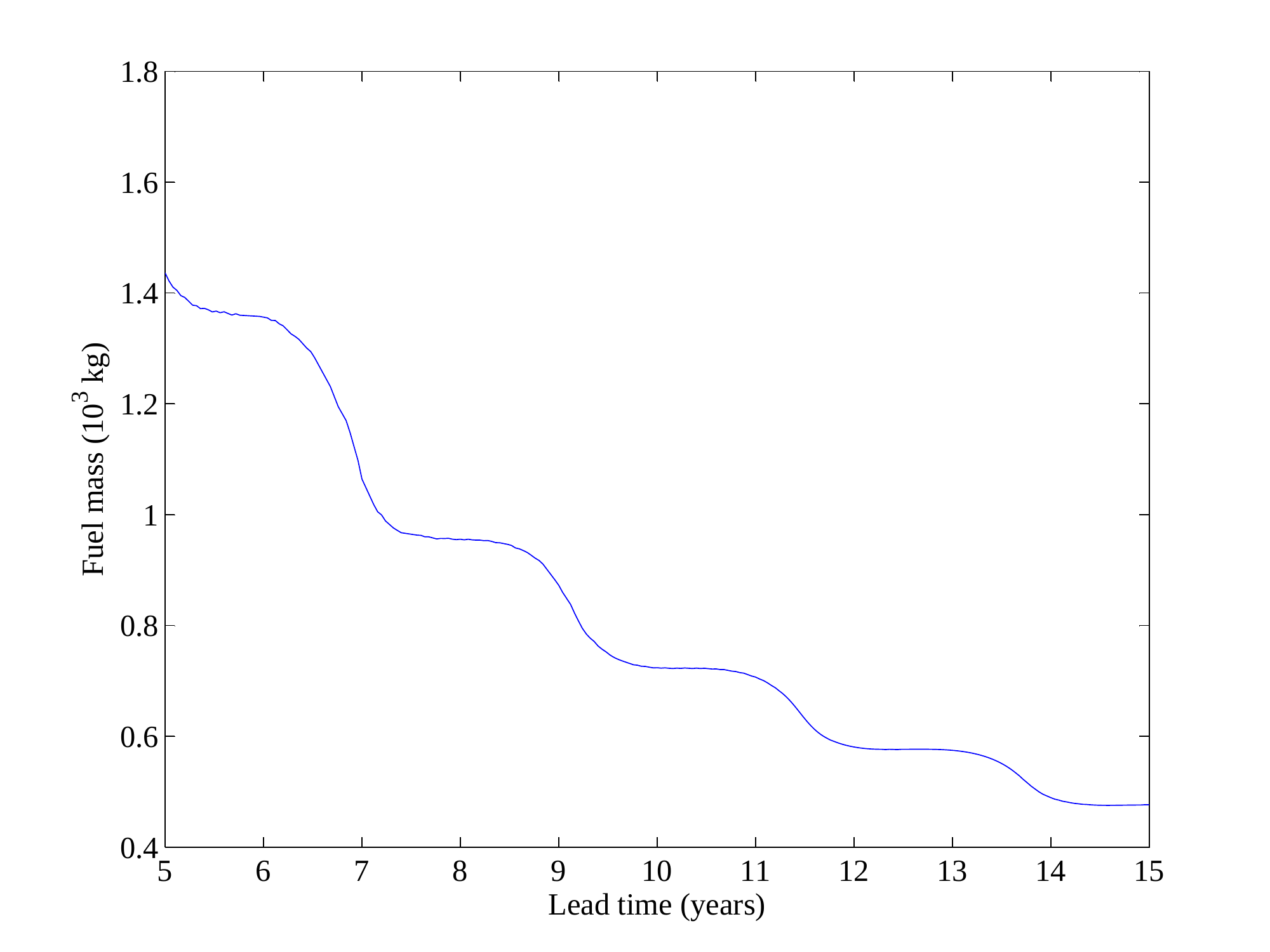}}
    \end{picture}
  \end{center}
  \caption{Optimal fuel mass burned for the gravity tractor spacecraft as a
    function of lead time of start of deflection action.}
\label{fuel}
\end{figure}

Lastly Figure (\ref{fuel}) shows the dependence of the amount of fuel
that is required as a function of the lead time.
 
\section{Conclusions}
As a potential method of asteroid impact aversion, the gravity tractor has 
the advantage of not requiring any direct contact of the spacecraft with the
asteroid and therefore avoids problems related to the asteroid's
make-up and structural and material properties. The method relies on the relatively small-magnitude
gravitational force that can be exerted on the asteroid by a typically sized
spacecraft. 

The work presented in this paper describes an approach to increase the
efficiency of the gravity tractor by optimizing the amount of mass
needed to deflect an asteroid by a given amount. Thus, a
decrease of gross spacecraft mass for the same amount of deflection
has been obtained that is in the order of 20\%. This is a
decrease in mass that significantly improves the practicality of the
gravity tractor as an approach to asteroid deflection.
Thus, for an asteroid of a ``smaller size'' with a lead time in the
order of 5 to 15 years, the gravity tractor on a restricted Keplerian
orbit may be a viable option for asteroid deflection.

A limitation of the current study (and other studies regarding the use
of gravity tractors) is that it is assumed that the asteroid has a
spherically symmetric gravitational field. This is generally not the
case, and may require additional asteroid-centric navigation in order
to remain in a controlled orbit about the asteroid. This problem is
the subject of current studies.


\bibliography{main.bib}

\end{document}